\documentclass[10pt,aps,prc,amsmath,amssymb,twocolumn,showpacs,floatfix,byrevtex]{revtex4-1}
\usepackage{hyperref} 
\usepackage{graphicx,bm,bbm,xcolor,dcolumn}
\usepackage[12hr]{datetime}
\usepackage{paralist}
\usepackage{subcaption}
\usepackage{multirow}

\newcommand{\ChiEFT}{\ensuremath{\chi\mathrm{EFT}}}
\newcommand{\MeV}{\, \mbox{MeV}}

\newcommand{\NxLO}[1][3]{\ensuremath{\textrm{N}^{#1}\textrm{LO}}}

\newcommand{\pin}{\ensuremath{\pi{}N}}
\newcommand{\nn}{\ensuremath{NN}}

\newcommand{\cv}{\bm{c}}

\newcommand{\chired}{\ensuremath{\chi^2_{\mathrm{red}}}}
\newcommand{\WI}{WI08}

\renewcommand{\section}[1]{\textit{#1.}--}

\begin{document}
 
\title{Uncertainty Quantification of the Pion-Nucleon Low-Energy Coupling Constants up to Fourth Order in Chiral Perturbation Theory}

\author{K.~A.~Wendt} 
\email{kwendt2@utk.edu}
\affiliation{Department of Physics and Astronomy, University of Tennessee, Knoxville, Tennessee 37996, USA} 
\affiliation{Physics Division, Oak Ridge National Laboratory, Oak Ridge, Tennessee 37831, USA}

\author{B.~D.~Carlsson} 
\affiliation{Department of Fundamental Physics, Chalmers University of Technology, SE-412 96 G\"{o}teborg, Sweden}

\author{A.~Ekstr\"om}
\affiliation{Department of Physics and Astronomy, University of Tennessee, Knoxville, Tennessee 37996, USA} 
\affiliation{Physics Division, Oak Ridge National Laboratory, Oak Ridge, Tennessee 37831, USA}

\date{\currenttime~\today}

\begin{abstract}
We extract the statistical uncertainties for the pion-nucleon ($\pi
N$) low energy constants (LECs) up to fourth order
$\mathcal{O}(Q^4)$ in the chiral expansion of the nuclear effective
Lagrangian. The LECs are optimized with respect to experimental
scattering data. For comparison, we also present an uncertainty
quantification that is based solely on \pin{} scattering
phase shifts. Statistical errors on the LECs are critical in order
to estimate the subsequent uncertainties in \textit{ab initio}
modeling of light and medium mass nuclei which exploit chiral
effective field theory. As an example of the this, we present the
first complete predictions with uncertainty quantification of
peripheral phase shifts of elastic proton-neutron scattering.
\end{abstract}
\maketitle

\section{Introduction}
Chiral effective field theory (\ChiEFT{}) for nuclear physics provides
a theoretical framework for a common description of various nuclear
processes through the systematic generation of two- three- and many-
body interactions and currents.  Undoubtedly, it has allowed advances
in ab initio structure calculations of light~\cite{Maris2013} and
medium mass~\cite{Roth2012,Hergert2013,Hagen2012} atomic nuclei. The
quantitative predictions of \ChiEFT{} depend on the numerical values
of a set of low-energy constants (LECs), which have become a limiting
factor in medium mass nuclei where current sets of LECs fail to
simultaneously predict binding, spectra, and radii.  It is therefore
relevant to constrain these  LECs such that all predictions within the
realm of applicability of \ChiEFT{} can be quantified together with
statistical uncertainties. This is important for advancing modern
many-body calculations into regions of the nuclear chart and the
physical processes where experimental data for verification is
limited, such as neutrino-less double beta decay or
structure and reactions near the neutron drip line.  We present
results that constrain the \pin{}-sector of \ChiEFT{}, with
accompanying confidence intervals (CI), up to fourth order in the
expansion of the effective Lagrangian. At this order, the sub-leading
long range three-nucleon interaction enters, and will thus be fully
constrained by the results presented here.  Further, many operator
currents, such as the axial-vector current, depend on only \pin{} LECs
and three nucleon contact LECs. Therefore careful uncertainty
quantification of the \pin{} LECs is also critical for comparing
processes driven by these currents to experimental cross sections and
decay measurements.

The \ChiEFT{} interaction Lagrangian $\mathcal{L}_{\rm eff}$ for
atomic nuclei can be separated into two terms $\mathcal{L}_{\rm
eff}=\mathcal{L}_{\nn{}}+\mathcal{L}_{\pin{}}$, and the two different
contributions each depend explicitly on a distinctive set of LECs. The
first term parametrizes the short-ranged contact-interactions and the
second term describes the long-ranged and pion-mediated part of the
two- and three-nucleon interaction. While the \nn{}-contact sector
must be constrained using nucleon-nucleon data, the LECs in
$\mathcal{L}_{\pin{}}$ can be determined from experimental \pin{}
scattering-data, completely separately from the \nn{} terms. This is
one example of how \ChiEFT{} can link separate physical processes that
are relevant for the description of atomic nuclei. Previous
constraints for the \pin{} LECs have been determined from peripheral
\nn{} scattering phase shifts~\cite{Entem2002,Ekstrom2013} or \pin{}
elastic scattering phase
shifts~\cite{Buttiker2000,Fettes2000a,Krebs2012}. Indeed, these
efforts have produced various sets of LECs that closely reproduce the
respective phase shift analyses (either
\nn{}~\cite{Stoks1993,Arndt2007} or
\pin{}~\cite{Koch1985,Koch1986,Workman2012,SAID:WI08});
however, the lack of reliable uncertainties on the input phase shifts
prevents meaningful uncertainty quantification of the \pin{} LECs. It
should be noted that available scattering phase shifts are not
experientially measured data, but the result of a partial-wave
analysis of measured data. In contrast with previous determinations of
the \pin{} LECs, the analysis presented here is grounded in
experimental scattering data. This allows us to estimate meaningful
statistical uncertainties. For the first time we can therefore explore
the consistency of \ChiEFT{} by predicting CI's for the peripheral
\nn{} phase shifts determined by \pin{}-data.

\begin{table}
  \caption{\label{tab:lecs}Numerical values of the \pin{} LECs that
  result from the optimization with respect to explrimental
  observables. The resulting values are grouped from left to right in
  the order they appear in the Lagrangian.}
  \begin{ruledtabular}
  \begin{tabular}{r D{,}{\,\pm\,}{-1} | r D{,}{\,\pm\,}{-1} | r D{,}{\,\pm\,}{-1}} 
     \multicolumn{2}{c|}{$\mathcal{O}(Q^1)$ LECs } & \multicolumn{2}{c|}{$\mathcal{O}(Q^2)$ LECs} & \multicolumn{2}{c}{$\mathcal{O}(Q^3)$ LECs}\\
     \multicolumn{2}{c|}{[GeV$^{-1}$]} & \multicolumn{2}{c|}{[GeV$^{-2}$]} & \multicolumn{2}{c}{[GeV$^{-3}$]} \\ 
     \hline
      $c_{1}$ &  -1.40,0.12 &       $\bar{d}_1+\bar{d}_2$ &  +5.80,0.14 &$\bar{e}_{14}$ &  +1.53,0.31\\
      $c_{2}$ &  +1.71,0.33 &                 $\bar{d}_3$ &  -5.66,0.08 &$\bar{e}_{15}$ & -11.91,0.87\\
      $c_{3}$ &  -4.56,0.11 &                 $\bar{d}_5$ &  +0.03,0.06 &$\bar{e}_{16}$ & +11.43,1.23\\
      $c_{4}$ &  +3.72,0.27 & $\bar{d}_{14}-\bar{d}_{15}$ & -11.50,0.12 &$\bar{e}_{17}$ &  +0.73,0.51\\
              &             &                             &             &$\bar{e}_{18}$ &  +0.57,1.36\\
  \end{tabular}
  \end{ruledtabular}
\end{table}

\section{Optimization}
We seek a set of \pin{} LECs $\cv_{\star}$ that minimize the least-squares objective
function
\footnote{See Refs.~\cite{Arndt2007} for details on their process for
a slightly older solution (SP06)}:
\begin{align}
  \label{eqn:Chi2:Obs}
  \chired(\cv,\bm{N}) &= \frac{1}{n_{df}}\left(\sum_{i}R_{i}(\cv,\bm{N})^2+\sum_j r_i(\bm{N})^2\right)
  \\
  R_i(\cv,\bm{N}) &= \frac{N_{j_i}O_{i}^{\ChiEFT{}}(\cv)-O_{i}^{\mathrm{Exp.}}}{\Delta_{i}^{Exp.}} 
  \\
  r_j(\bm{N}) &= \frac{N_j-1}{\Delta_j} 
\end{align}
where $O_{i}^{\ChiEFT{}}(\cv)$ denotes the value of the scattering
observable computed from \ChiEFT{}, while $O_{ji}^{\mathrm{Exp.}}$ and
$\Delta_{ji}^{Exp.}$ denotes the experimentally measured value and
uncertainty, respectively, for the corresponding observable. $\cv$ is
a vector of LECs spanning all included $c_i$, $d_i$, and $e_i$ LECs.
$\bm{N}$ is a vector of normalization coefficients $N_j$, where all
points from a single experimental angular distribution share the same
$N_j$; $\Delta_j$ encodes uncertainty of the experimental systematics
for $N_j$.  The number of degrees of freedom is given by $n_{\rm df} =
(n_d+n_{N}-n_{NF})-(n_{\rm lecs}+n_{N})$, where $n_d$ is the number of
data included in the fit, $n_{\rm lecs}$ is the number of LECs being
fit, and $n_N$ is the number of unknown normalization coefficients and
$n_{NF}$ is the number of floated coefficients (contribute no residual
term in $r_j$ as $\delta_j=\infty$). For a given experiment $j$, the
included data points run over a series of scattering angles at a fixed
lab frame momentum $Q_{\rm Lab}$.

For our fitting dataset, we adopt the database from the most recent 
\pin{} partial wave analysis~\cite{Workman2012}, referred to as 
\WI{}.  By construction, \ChiEFT{} is a low-energy theory, therefore
we exclude data with lab-frame momentum $Q_{\mathrm{Lab}}>160\MeV$.
This leaves us with an experimental database consisting of
differential scattering cross-sections and polarization cross sections
from $\pi^{\pm}+p\rightarrow\pi^{\pm}+p$ and
$\pi^{-}+p\rightarrow\pi^{0}+n$ processes. In total, there are
$n_d=1246$ data points, consisting of $1194$ differential unpolarized
cross-sections and $52$ differential singly-polarized cross-sections.
There are $n_N=110$ normalization coefficients, with $n_{NF}=9$
floated coefficients.  There are other measurable observables,
such as the spin rotation parameters, but experimental data only
exists for momentum well beyond the range of validity of the EFT.  The
cutoff in lab frame momentum ($160\MeV$) was chosen such that
increasing or decreasing the cutoff would lead to a larger minimum
value of $\chired(\cv,\bm{N})$, maximizing amount of included data
while avoiding fitting past the radius of convergence of the EFT.

For the calculated observables, we use the strong amplitudes presented
in Ref.~\cite{Krebs2012}
(Refs.~\cite{Fettes1998,Fettes2000,Fettes2000a,Fettes2001} give a more
complete presentation of the \pin{} scattering amplitudes, but use a
different power counting scheme for relativistic corrections.) For
the strong amplitude, we adopt their conventions for fixing
$\bar{d}_{18}$, absorbing $\bar{e}_{19,20,21,22,35,36,37,38}$,
$\bar{l}_{3}$ into $c_{1,2,3,4}$.  We also work with exact isospin
symmetry and use an averaged pion mass ($m_{\pi^{\vphantom{\pm}}} =
(m_{\pi^0}+2m_{\pi^\pm})/3$). We adopt the electromagnetic treatment
that is used in the \WI{} partial wave analysis, which is described in
detail inRefs.~\cite{Tromborg1978,Tromborg1974,Tromborg1977,Bugg1973}.  For the
electromagnetic corrections, we explicitly break isospin symmetry and
use the physical pion masses within the coulomb amplitudes.  Actual
fits were performed using the TAO package~\cite{TAO}.
  
\section{Results}
The central values and $1\sigma$ uncertainties of the \pin{} LECs up
to fourth order in \ChiEFT{} from fitting against scattering data are
presented in Table~\ref{tab:lecs}. For this fit, we find that
$\chired=2.29$.  As a comparison, the LECs from the fit against  \WI{}
phase shifts of Ref.~\cite{Krebs2012} generate $\chired=3.63$  with
respect to our objective function.  Not surprisingly, our fit will
reproduce experimental data better than the fits with respect to phase
shifts. While our LECc are consistent with the spread of previous
analyses, we find that no single analysis is entirely consistent with
our fit at the $95\%$  confidence level, though the WI08 \pin{} phase
shift fit from Ref.~\cite{Krebs2012} lies just outside our interval.

\begin{table*}
  \caption{\label{tab:covcor}Covariance (lower triangle) and
  correlation (upper triangle) matrices for our fit to experimental
  data}
  \begin{ruledtabular}
  \begin{tabular}{c| D{.}{.}{-1} D{.}{.}{-1} D{.}{.}{-1} D{.}{.}{-1} D{.}{.}{-1} D{.}{.}{-1} D{.}{.}{-1} D{.}{.}{-1} D{.}{.}{-1} D{.}{.}{-1} D{.}{.}{-1} D{.}{.}{-1} D{.}{.}{-1}  } 
      LEC & \multicolumn{1}{c}{$c_{1}$} & \multicolumn{1}{c}{$c_{2}$} & \multicolumn{1}{c}{$c_{3}$} & \multicolumn{1}{c}{$c_{4}$} & \multicolumn{1}{c}{$\bar{d}_1+\bar{d}_2$} & \multicolumn{1}{c}{$\bar{d}_3$} & \multicolumn{1}{c}{$\bar{d}_5$} & \multicolumn{1}{c}{$\bar{d}_{14}-\bar{d}_{15}$} & \multicolumn{1}{c}{$\bar{e}_{14}$} & \multicolumn{1}{c}{$\bar{e}_{15}$} & \multicolumn{1}{c}{$\bar{e}_{16}$} & \multicolumn{1}{c}{$\bar{e}_{17}$} & \multicolumn{1}{c}{$\bar{e}_{18}$}\\
    \cline{1-14}
                       $c_{1}$ & +0.03 & \multicolumn{1}{|D{.}{.}{-1}}{+0.95} & +0.24 & +0.56 & +0.51 & -0.33 & -0.48 & -0.45 & -0.31 & +0.37 & -0.85 & +0.15 & -0.54\\
    \cline{3-3}
                       $c_{2}$ & +0.07 & +0.21 & \multicolumn{1}{|D{.}{.}{-1}}{-0.08} & +0.55 & +0.54 & -0.37 & -0.48 & -0.56 & -0.32 & +0.56 & -0.96 & +0.18 & -0.54\\
    \cline{4-4}
                       $c_{3}$ & +0.01 & -0.01 & +0.02 & \multicolumn{1}{|D{.}{.}{-1}}{+0.13} & -0.02 & +0.12 & -0.06 & +0.26 & +0.08 & -0.58 & +0.26 & -0.12 & -0.08\\
    \cline{5-5}
                       $c_{4}$ & +0.03 & +0.09 & +0.01 & +0.14 & \multicolumn{1}{|D{.}{.}{-1}}{+0.97} & -0.75 & -0.77 & -0.74 & -0.39 & +0.36 & -0.55 & +0.18 & -0.94\\
    \cline{6-6}
          $\bar{d}_1+\bar{d}_2$ & +0.01 & +0.05 & -0.00 & +0.07 & +0.03 & \multicolumn{1}{|D{.}{.}{-1}}{-0.81} & -0.76 & -0.75 & -0.43 & +0.46 & -0.57 & +0.18 & -0.91\\
    \cline{7-7}
                    $\bar{d}_3$ & -0.01 & -0.02 & +0.00 & -0.03 & -0.02 & +0.01 & \multicolumn{1}{|D{.}{.}{-1}}{+0.24} & +0.67 & +0.43 & -0.45 & +0.44 & -0.24 & +0.74\\
    \cline{8-8}
                    $\bar{d}_5$ & -0.01 & -0.02 & -0.00 & -0.02 & -0.01 & +0.00 & +0.00 & \multicolumn{1}{|D{.}{.}{-1}}{+0.52} & +0.25 & -0.28 & +0.48 & -0.04 & +0.69\\
    \cline{9-9}
    $\bar{d}_{14}-\bar{d}_{15}$ & -0.01 & -0.04 & +0.01 & -0.04 & -0.02 & +0.01 & +0.01 & +0.02 & \multicolumn{1}{|D{.}{.}{-1}}{+0.34} & -0.52 & +0.62 & -0.47 & +0.81\\
    \cline{10-10}
                 $\bar{e}_{14}$ & -0.02 & -0.05 & +0.00 & -0.05 & -0.03 & +0.02 & +0.01 & +0.02 & +0.13 & \multicolumn{1}{|D{.}{.}{-1}}{-0.80} & +0.50 & -0.28 & +0.41\\
    \cline{11-11}
                 $\bar{e}_{15}$ & +0.07 & +0.29 & -0.10 & +0.15 & +0.10 & -0.05 & -0.02 & -0.09 & -0.33 & +1.30 & \multicolumn{1}{|D{.}{.}{-1}}{-0.76} & +0.32 & -0.40\\
    \cline{12-12}
                 $\bar{e}_{16}$ & -0.22 & -0.71 & +0.06 & -0.33 & -0.17 & +0.07 & +0.05 & +0.15 & +0.30 & -1.40 & +2.59 & \multicolumn{1}{|D{.}{.}{-1}}{-0.25} & +0.56\\
    \cline{13-13}
                 $\bar{e}_{17}$ & +0.01 & +0.05 & -0.01 & +0.04 & +0.02 & -0.02 & -0.00 & -0.04 & -0.06 & +0.22 & -0.24 & +0.36 & \multicolumn{1}{|D{.}{.}{-1}}{-0.50}\\
    \cline{14-14}
                 $\bar{e}_{18}$ & -0.14 & -0.41 & -0.02 & -0.58 & -0.27 & +0.13 & +0.08 & +0.21 & +0.24 & -0.75 & +1.48 & -0.49 & +2.70
  \end{tabular}
  \end{ruledtabular}
\end{table*}

For uncertainty analysis, we apply a standard gradient expansion of
$\chi^2(\cv)$  at the optimum $\cv_{\star}$(see
Ref.~\cite{Dobaczewski2014} and references therein for further detail):
\begin{equation}
  \chi^2(\cv) = \chi^2(\cv_{\star}) +
      \frac{1}{2}\sum_{a,b} \left(\cv-\cv_{\star}\right)_a  \bm{H}_{a,b} 
      \left(\cv-\cv_{\star}\right)_b +... \,,
\end{equation}
\begin{equation}
      \bm{H}_{a,b} = \frac{\partial^2\chi^2}{\partial c_a \partial c_a} \bigg|_{\cv=\cv_{\star}}\,,
\end{equation}
where $\chi^2(\cv)$ is the full (not reduced by $n_{df}$) objective
function. If the residual vectors ($R_i$ and $r_j$) are normally
distributed, then the covariance matrix of our fit is given by
\begin{align}
  \label{eqn:cov}
  \bm{C}_{a,b} = \operatorname{cov}(c_ac_b) &\approx \chired \operatorname{inv}(\bm{H})_{a,b}\,,
  \\
  \operatorname{corr}(c_ac_b) &= \bm{C}_{a,b}/\sqrt{\bm{C}_{a,a}\bm{C}_{b,b}}\,.
\end{align}.

The covariance matrix ($\bm{C}_{a,b}$) and correlation matrix
$\operatorname{corr}(c_ac_b)$  are presented in
table~\ref{tab:covcor}.

\begin{figure*} 
  \centering
  \includegraphics{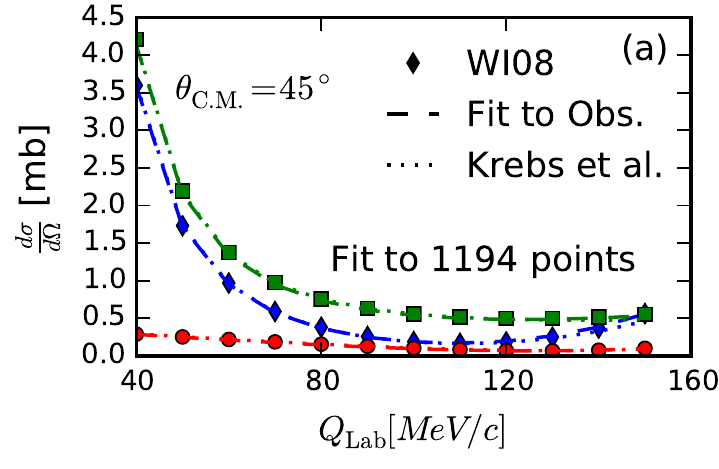}
  \includegraphics{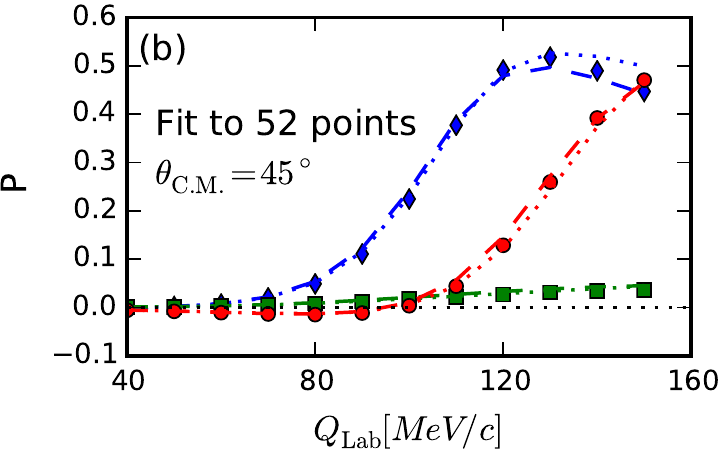}
  \\
  \includegraphics{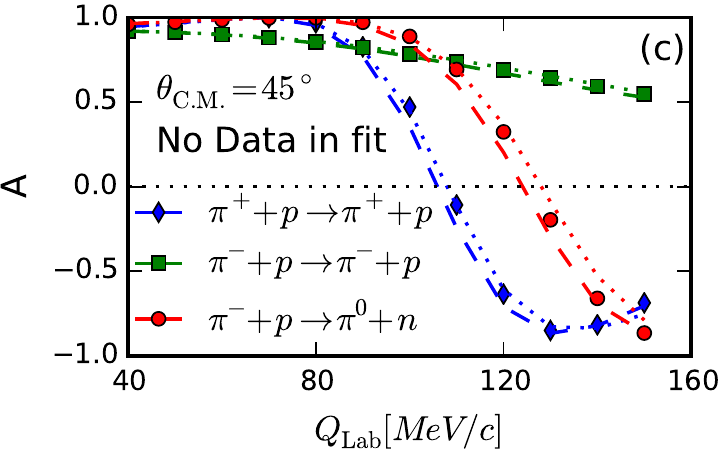}
  \includegraphics{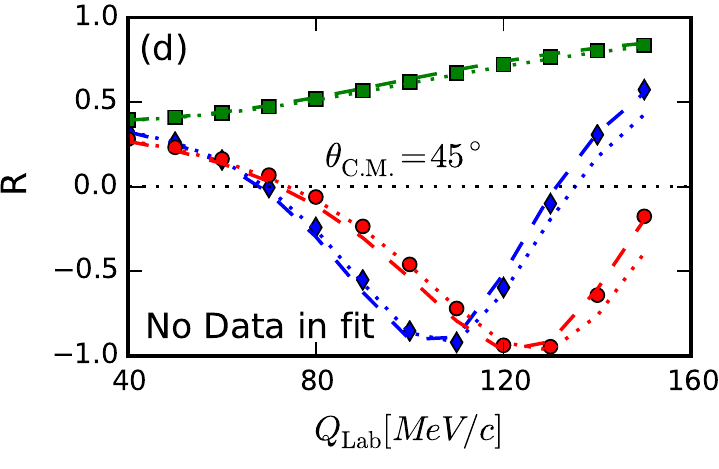}
  \caption{\pin{} Observables computed at $\theta_{\rm
  c.m.}=45^\circ$. The blue (green) lines with diamond (square)
  markers show results for the
  $\pi{}^{+(-)}+p\rightarrow\pi{}^{+(-)}+p$ processes while the red
  lines with circular markers show the charge exchange process.  The
  dashed lines show observables computed using our fit to experimental
  data.  The dotted lines shows results reconstructed from  LECs of
  Ref.~\cite{Krebs2012}.  P is the polarization, while A and R are the
  spin rotation parameters, all presented as ratios with respect to
  the differential cross-section.  Definitions of these observables
  can be found in Ref.~\cite{Hohler1983}
  \label{fig:PiN:Obs}}
\end{figure*}

Figure~\ref{fig:PiN:Obs} shows calculations of \pin{} scattering
observables using our LECs fit data as well as LECs from the phase
shift fit of Ref.~\cite{Krebs2012}.  At smaller $Q_{\rm lab}$, the
difference is minimal, but it is clear at higher momentum that phase
shifts fits are inadequate.  This is especially apparent in the
calculations of the polarization (P) and spin rotation parameters (A
and B), suggesting a phase shift fit may not adequately capture the
underlying tensor effects that are critical to nuclear observables.

\begin{figure*} 
  \centering
  \includegraphics{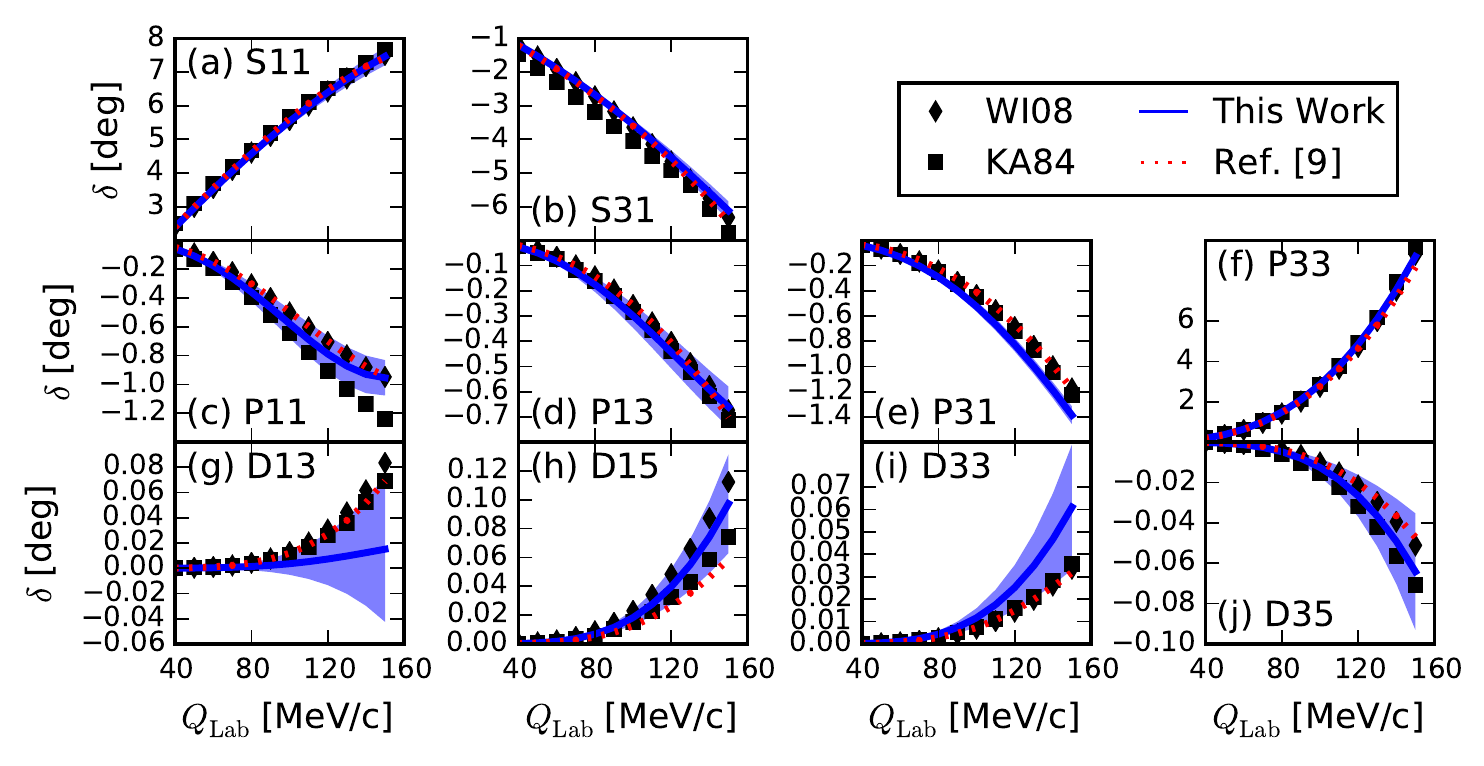}
  \caption{\pin{} phase shifts.  The blue line is a prediction of 
           our fit with the band showing the $95\%$ interval.  For
           comparison, the red dotted lines are the phase shifts from
           \cite{Krebs2012}. The markers show phase shifts from two
           partial wave analyses~\cite{Koch1985,Workman2012}.  The
           partial waves are denoted as [L][2I][2J] where L is the
           orbital angular momentum,  I is the isospin, and J is the
           total angular momentum.
           \label{fig:PiN:PhaseShifts}}
  
\end{figure*}

As a validation of our analysis, we plot \pin{} partial wave phase
shifts in Fig.~\ref{fig:PiN:PhaseShifts} and compare with two
different partial wave analyses(\WI{}~\cite{Workman2012} and
KA84~\cite{Koch1985}) The blue bands denotes the $95\%$ CI from our
fit.  For S- and P-waves, we have good agreement with the \WI{}
partial wave analysis.  In the D-waves, where the $\mathcal{O}(Q^3)$
LECs have significant contributions, we find poor agreement in
the $J=\tfrac{3}{2}$ channels (D13 and D33).

To better understand where this error in the $J=\tfrac{3}{2}$ D-waves
is coming from, we can use the eigenstate of the covariance matrix to
gain insight on the quality of the LEC constraints:
\begin{align}
  \bm{C} \bm{v}_j &= \nu_j \bm{v}_j.
\end{align}
The vectors of LEC combinations, $\bm{v}_j$, form pseudo LECs that are
completely uncorrelated, and eigenvalues $\nu_j$ are effective
variances of these pseudo LECs.  In table~\ref{tab:pca}, we examine
the LEC content of the least constrained eigenvectors (those with
largest eigenvalue). It is clear that the largest eigenvectors are
dominated by the $\bar{e}_i$ LECs, and  in particular the pair
$\bar{e}_{16}$, $\bar{e}_{18}$ which dominate  the two most
significant vectors. This indicates that $\bar{e}_{16}$ and
$\bar{e}_{18}$ are poorly constrained.  Within the scattering
amplitudes, $\bar{e}_{16}$ and $\bar{e}_{18}$ span the non-spin-flip
and spin-flip amplitudes respectively~\cite{Krebs2012},  suggesting
that a lack of low momentum data for spin observables could be at
fault.

  \begin{table}
    \caption{\label{tab:pca} Eigenvalues ($\nu_j$) of
      the covariance matrix, with contributions to eigenvector ($v_j$)
      summed over LECs of same chiral order.  These eigenvalue
      correspond to the variances $\sigma^2_j$ of effective
      uncorrelated pseudo LECs. For this analysis the covariance
      matrix was reduced to a dimensionless form using appropriate
      powers of the nucleon mass.}
    \begin{ruledtabular}
    \begin{tabular}{rrrrrrr} 
      \multicolumn{1}{c}{j} &
      \multicolumn{1}{c}{$\nu_j$} &
      \multicolumn{1}{c}{$\sum_{c_i}v^2_{j,c_i}$} &
      \multicolumn{1}{c}{$\sum_{\bar{d}_i}v^2_{j,\bar{d}_i}$} &
      \multicolumn{1}{c}{$\sum_{\bar{e}_i}v^2_{j,\bar{e}_i}$} &
      \multicolumn{1}{c}{$v^2_{j,\bar{e}_{16}}$} &
      \multicolumn{1}{c}{$v^2_{j,\bar{e}_{18}}$} \\
      \hline
       1 & 4.1147 & 0.0525 & 0.0094 & 0.9382 & 0.3829 & 0.4207  \\
       2 & 1.4185 & 0.0362 & 0.0052 & 0.9587 & 0.2558 & 0.4798  \\
       3 & 0.5114 & 0.1067 & 0.0013 & 0.8920 & 0.2038 & 0.0038  \\
       4 & 0.2412 & 0.0748 & 0.0236 & 0.9016 & 0.0241 & 0.0157  \\
       5 & 0.0516 & 0.2721 & 0.0459 & 0.6819 & 0.0004 & 0.0021  \\
       6 & 0.0060 & 0.0076 & 0.9696 & 0.0227 & 0.0001 & 0.0044  \\
       7 & 0.0047 & 0.0134 & 0.9647 & 0.0220 & 0.0006 & 0.0008  \\
       8 & 0.0015 & 0.2194 & 0.6919 & 0.0887 & 0.0147 & 0.0368  \\
       9 & 0.0007 & 0.6323 & 0.2427 & 0.1250 & 0.0790 & 0.0002  \\
      10 & 0.0003 & 0.8666 & 0.0346 & 0.0987 & 0.0235 & 0.0348  \\
      11 & 0.0001 & 0.7617 & 0.0125 & 0.2258 & 0.0007 & 0.0007  \\
      12 & 0.0000 & 0.0080 & 0.9918 & 0.0002 & 0.0000 & 0.0000  \\
      13 & 0.0000 & 0.9488 & 0.0068 & 0.0444 & 0.0144 & 0.0001  \\
    \end{tabular}
    \end{ruledtabular}
  \end{table}

\section{Application to Uncertainty of \nn{} Phase Shifts}
Accurate calculations accompanied with quantitative uncertainty
estimates is of-course one of the principal goals of low-energy
nuclear theory. This is critical to the study of nuclei near the
neutron drip-line or for $0\nu\beta\beta$ decay where this
insufficient data to validate many-body calculations.

As a first step towards quantifying the accuracy of the \pin{} sector
of the \nn{} interaction, we will predict the confidence intervals of
the peripheral ($J\ge4$) \nn{} scattering phase shifts at \NxLO[3]{}. These
partial waves are fully determined by the long-ranged pion physics. It
should also be noted that this is the first \NxLO[3]{} calculation of these
phase shifts that is actually grounded in experimental scattering
data.

\begin{figure*}
  \includegraphics{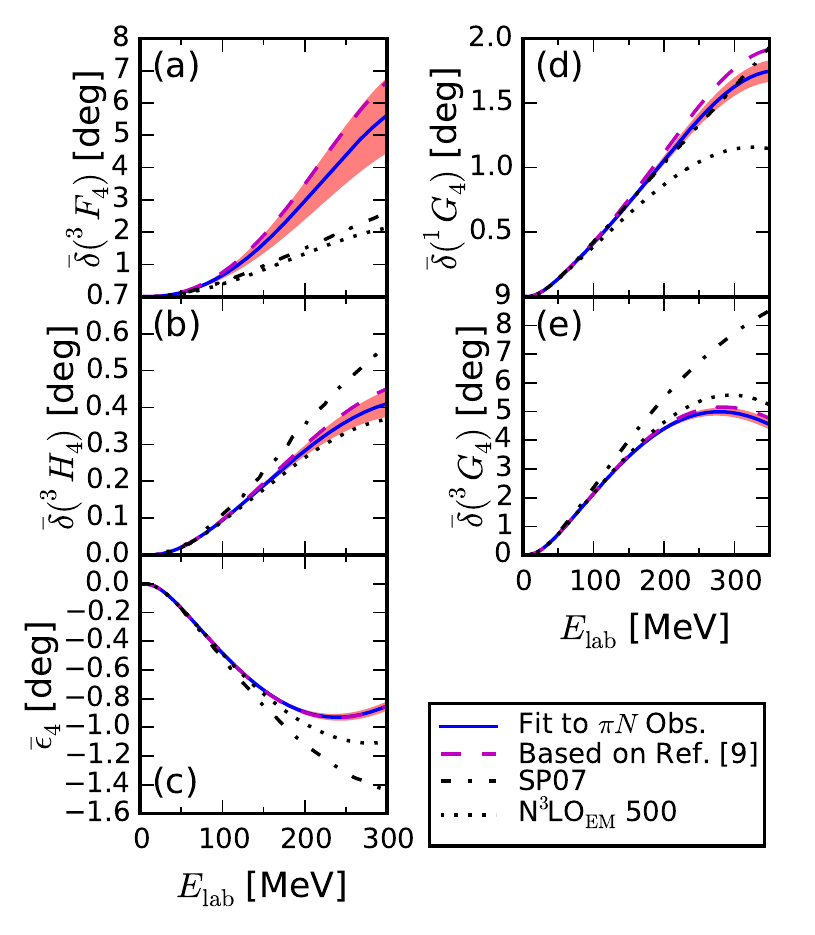}
  \includegraphics{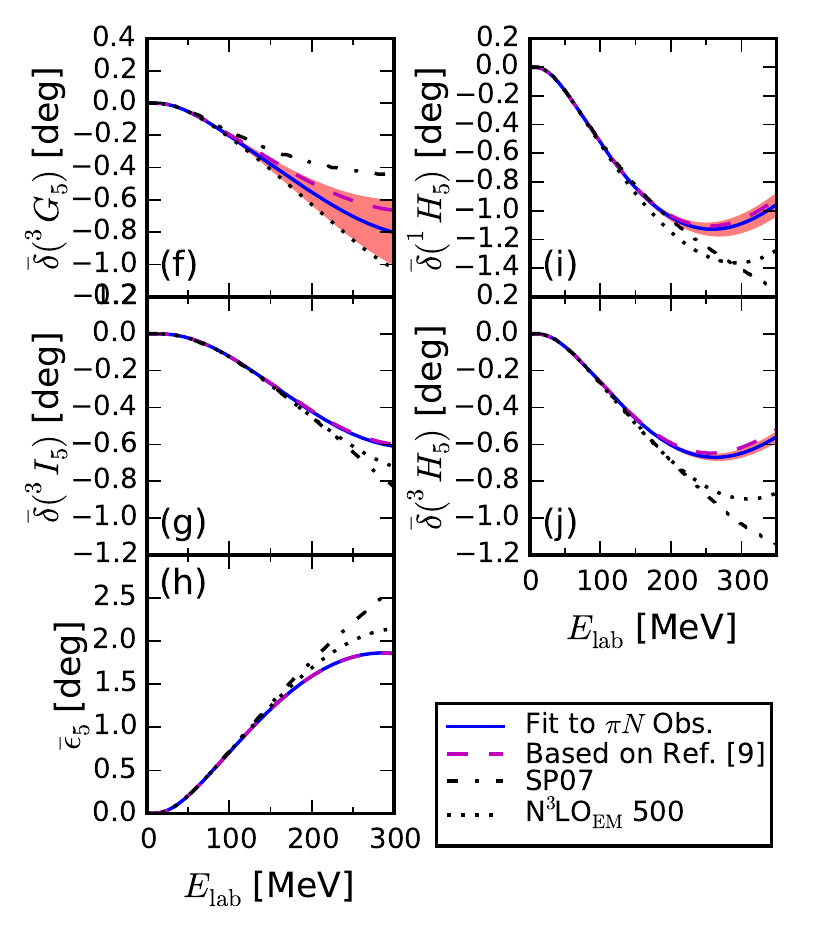}
  \caption{Elastic proton-neutron $J=4,5$ phase shifts at \NxLO[3]{}
    ($\Lambda=500$ MeV) with $95\%$ CIs from \pin{} data (red band).
    (Dotted line) the phase shifts from the \NxLO[3]{} interactions
    of Ref.~\cite{Entem2003}. (Dashed line) the phase shifts from the
    Nijmegen partial wave analysis of Ref.~\cite{Stoks1993}.}
  \label{fig:UQnn}
\end{figure*}

Figure~\ref{fig:UQnn} shows the predicted $95\%$ CIs of the proton-
neutron elastic scattering phase shifts for all total angular momentum
$J=4,5$ partial waves.  For comparison, phase shifts from two
different partial wave analyses are presented, PWA93~\cite{Stoks1993}
and SP07~\cite{Arndt2007}. We also compare to phase shifts computed
using the high-precision Idaho-\NxLO[3]{} interaction~\cite{Entem2002},
which reproduces the SM99~\cite{Machleidt2001} \nn{} database with
$\chired\sim{}1$, and to
\NxLO[3]{} \nn{} interactions computed using the LECs from
Ref.~\cite{Krebs2012}.  Our LECs  perform quite favorably
compared the \pin{} phase shift fit, and surprisingly favorable in
many channels to the phase shifts computed from Idaho-\NxLO[3]{}.  For many
channels, our error bands are remarkably small.

\section{Conclusions}
We constrained the \pin{} LECs using experimental data with
$\chi^2/\textrm{datum} = 2.29$, generating a set of LECs with error
estimates that are compatible for use with modern \NxLO{3}
Hamiltonians and currents. We find that our lower order LECs are of
natural size, while the higher order LECs tend to be unnaturally large (or
small in the case of $\bar{d}_5$.)  We find that even with fairly
large error bars for some LECs, the proton-neutron peripheral phase shifts
are very well constrained at lab scattering energies below $100\MeV$.
The \pin{} LECs not only fix the long range part of the \NxLO[3]{}
three body force, but our analysis provides a means to examine
uncertainty of the three body force in nuclear bound states as well as
uncertainty from currents in observables. Progressing forward,
simultaneous constraints of \pin{} and \nn{} LECs with quantitative
statical analysis will yield predictions of few- and many-body systems
with quantified uncertainty and is a topic for many future
investigations.

\section{Acknowledgments}
The authors would like to thank H.~Krebs, T.~Papenbrock, D.~Phillips,
and R.~Workman for their useful discussion.  This work was supported
in part by the U.S. Department of Energy (DOE) under Grant
Nos.~DEFG02-96ER40963 (University of Tennessee), DE-SC0008499 (NUCLEI
SciDAC collaboration), Oak Ridge National Laboratory the Research
Council of Norway under contract ISP-Fysikk/216699, and by the
European Research Council under the European Community's Seventh
Framework Programme (FP7/2007-2013) ERC grant agreement no. 240603.
Oak Ridge National Laboratory is supported by the DOE Office of
Science under Contract No.~DE- AC05-00OR22725.

%

\end{document}